\documentclass[aps,preprint, floatfix]{revtex4}

\usepackage{graphicx,color}
\usepackage{psfrag}
\usepackage{amssymb}
\usepackage{eso-pic}
\usepackage[dvips,letterpaper]{geometry} 
\usepackage{amsmath}

\begin{document}

%
%
%
%
\def\oti{{\otimes}}
\def\lb{ \left[ }
\def\rb{ \right]  }
\def\tilde{\widetilde}
\def\bar{\overline}
\def\hat{\widehat}
\def\*{\star}
\def\[{\left[}
\def\]{\right]}
\def\({\left(}		\def\BL{\Bigr(}
\def\){\right)}		\def\BR{\Bigr)}
	\def\BBL{\lb}
	\def\BBR{\rb}
%
%
\def\zb{{\bar{z} }}
\def\zbar{{\bar{z} }}
\def\frac#1#2{{#1 \over #2}}
\def\inv#1{{1 \over #1}}
\def\half{{1 \over 2}}
\def\d{\partial}
\def\der#1{{\partial \over \partial #1}}
\def\dd#1#2{{\partial #1 \over \partial #2}}
\def\vev#1{\langle #1 \rangle}
\def\ket#1{ | #1 \rangle}
\def\rvac{\hbox{$\vert 0\rangle$}}
\def\lvac{\hbox{$\langle 0 \vert $}}
\def\2pi{\hbox{$2\pi i$}}
\def\e#1{{\rm e}^{^{\textstyle #1}}}
\def\grad#1{\,\nabla\!_{{#1}}\,}
\def\dsl{\raise.15ex\hbox{/}\kern-.57em\partial}
\def\Dsl{\,\raise.15ex\hbox{/}\mkern-.13.5mu D}
%
%
\def\ga{\gamma}		\def\Ga{\Gamma}
\def\be{\beta}
\def\al{\alpha}
\def\ep{\epsilon}
\def\vep{\varepsilon}
\def\la{\lambda}	\def\La{\Lambda}
\def\de{\delta}		\def\De{\Delta}
\def\om{\omega}		\def\Om{\Omega}
\def\sig{\sigma}	\def\Sig{\Sigma}
\def\vphi{\varphi}

%
%
\def\CA{{\cal A}}	\def\CB{{\cal B}}	\def\CC{{\cal C}}
\def\CD{{\cal D}}	\def\CE{{\cal E}}	\def\CF{{\cal F}}
\def\CG{{\cal G}}	\def\CH{{\cal H}}	\def\CI{{\cal J}}
\def\CJ{{\cal J}}          \def\CK{{\cal K}}	\def\CL{{\cal L}}
\def\CM{{\cal M}}	\def\CN{{\cal N}}	\def\CO{{\cal O}}
\def\CP{{\cal P}}	\def\CQ{{\cal Q}}	\def\CR{{\cal R}}
\def\CS{{\cal S}}	\def\CT{{\cal T}}	\def\CU{{\cal U}}
\def\CV{{\cal V}}	\def\CW{{\cal W}}	\def\CX{{\cal X}}
\def\CY{{\cal Y}}	\def\CZ{{\cal Z}}

\def\rvac{\hbox{$\vert 0\rangle$}}
\def\lvac{\hbox{$\langle 0 \vert $}}
\def\comm#1#2{ \BBL\ #1\ ,\ #2 \BBR }
\def\2pi{\hbox{$2\pi i$}}
\def\e#1{{\rm e}^{^{\textstyle #1}}}
\def\grad#1{\,\nabla\!_{{#1}}\,}
\def\dsl{\raise.15ex\hbox{/}\kern-.57em\partial}
\def\Dsl{\,\raise.15ex\hbox{/}\mkern-.13.5mu D}
%
%
%
\font\numbers=cmss12
\font\upright=cmu10 scaled\magstep1
\def\stroke{\vrule height8pt width0.4pt depth-0.1pt}
\def\topfleck{\vrule height8pt width0.5pt depth-5.9pt}
\def\botfleck{\vrule height2pt width0.5pt depth0.1pt}
\def\Zmath{\vcenter{\hbox{\numbers\rlap{\rlap{Z}\kern
0.8pt\topfleck}\kern 2.2pt
                   \rlap Z\kern 6pt\botfleck\kern 1pt}}}
\def\Qmath{\vcenter{\hbox{\upright\rlap{\rlap{Q}\kern
                   3.8pt\stroke}\phantom{Q}}}}
\def\Nmath{\vcenter{\hbox{\upright\rlap{I}\kern 1.7pt N}}}
\def\Cmath{\vcenter{\hbox{\upright\rlap{\rlap{C}\kern
                   3.8pt\stroke}\phantom{C}}}}
\def\Rmath{\vcenter{\hbox{\upright\rlap{I}\kern 1.7pt R}}}
\def\Z{\ifmmode\Zmath\else$\Zmath$\fi}
\def\Q{\ifmmode\Qmath\else$\Qmath$\fi}
\def\N{\ifmmode\Nmath\else$\Nmath$\fi}
\def\C{\ifmmode\Cmath\else$\Cmath$\fi}
\def\R{\ifmmode\Rmath\else$\Rmath$\fi}

\def\barray{\begin{eqnarray}}
\def\earray{\end{eqnarray}}
\def\beq{\begin{equation}}
\def\eeq{\end{equation}}

\def\n{\noindent}

\def\Tr{\rm Tr} 
\def\xvec{{\bf x}}
\def\kvec{{\bf k}}
\def\kvecp{{\bf k'}}
\def\omk{\om_{\kvec}} 
\def\2pi2{(2\pi)^2}
\def\ket#1{|#1 \rangle}
\def\bra#1{\langle #1 |}
\def\adag{a^\dagger}
\def\rme{{\rm e}}
\def\Im{{\rm Im}}
\def\pvec{{\bf p}}
\def\fermiS{\CS_F}
\def\cdag{c^\dagger}
\def\adag{a^\dagger}
\def\bdag{b^\dagger}
\def\vvec{{\bf v}}
\def\vac{|0\rangle}
\def\gradvec{\vec{\nabla}}
\def\psidag{\psi^\dagger} 
\def\up{\uparrow}
\def\down{\downarrow}
\def\smallhalf{{\textstyle \inv{2}}}
\def\smallsqrt{{\textstyle \inv{\sqrt{2}}}}
\def\rvec{{\bf r}}
\def\avec{{\bf a}}
\def\pivec{{\vec{\pi}}}
\def\dim#1{\lbrack\!\lbrack #1 \rbrack\! \rbrack }
\def\angstrom{{{\scriptstyle \circ} \atop A}     }
\def\AA{\leavevmode\setbox0=\hbox{h}\dimen0=\ht0 \advance\dimen0 by-1ex\rlap{
\raise.67\dimen0\hbox{\char'27}}A}

\def\Im{{\rm Im}}
\def\Re{{\rm Re}}

\def\Li{{\rm Li}}

\def\dim#1{{\rm dim}[#1]}

\def\ep{\epsilon}

\def\free{\CF}

\def\Fhat{\digamma}

\def\CI{\mathcal{I}}

\def\eF{\epsilon_F}

\def\pvec{{\bf p}}

\def\Kvec{{\bf K}}

\def\kappavec{\vec{\kappa}}

\def\ihat{\hat{i}}
\def\jhat{\hat{j}}

\def\cutoff{\delta \kappa}

\def\qvec{{\bf q}}

\def\vhat{\hat{v}}

\def\bfu{{\bf u}}

\def\bfM{{\bf M}}

\def\dk{ \frac{ d^2 \kvec }{(2\pi)^2 } }

\def\delS{ \delta S_F }

\def\delSF{\delS}

\def\Lex{\CL}

\def\Deltaxi{\Delta_\xi}

\def\Veffex{\CV_{\rm eff}^{(ex)}}

\def\Vex{\Veffex}

\def\Veffscreen{\CV_{\rm eff}^{(sc)}}

\def\qvec{{\bf{q}}}

\def\xihat{{\hat{\xi}}}

\def\Lsc{L^{(sc)}}
\def\Lex{L^{(ex)}}

\def\Lscreen{\Lsc}

\def\dpvec{  \frac{d^2 \pvec}{(2\pi)^2} }

\def\gsc{g}

\def\Deltagap{\Delta_{\rm gap}}

\def\smll{\scriptstyle}

\def\gR{g_{\rm sc}}

\title{Possible Cooper  instabilities in pair Green functions of  the two-dimensional  Hubbard model.}
\author{ Andr\'e  LeClair}
\affiliation{Newman Laboratory, Cornell University, Ithaca, NY}

\begin{abstract}

In analogy with ordinary BCS superconductivity,  we identify possible pairing instabilities with poles 
in a certain class of Green functions of the two-dimensional  repulsive Hubbard model,  which
normally signify bound states.  Relating the gap to the location of these poles,  we find that
these instabilities exist in the range of hole doping between $0.03$ and $0.24$.    The magnitude of
the gap can be calculated without introducing an explicit cut-off,   and for reasonably large
coupling $U/t = 10$,   the maximum gap is
on the order of $0.08$ in units of the hopping parameter $t$.   

\end{abstract}

\maketitle

\tableofcontents

\newpage

\section{Introduction}

The microscopic physics underlying  high $T_c$ superconductivity in the cuprates is
believed to be purely electronic in origin,  in contrast to ordinary superconductors where
the attractive mechanism is due to phonons.    Strongly correlated electron  models such
as the two-dimensional  Hubbard model have been proposed to describe it.  
The Hubbard model simply describes electrons hopping on a square lattice subject to 
strong, local, coulombic  {\it repulsion}.    Since it is known that the condensed charge carriers
have charge $2e$,  and thus some kind of Cooper pairing is involved,   the main open problem
is to identify the precise pairing mechanism.   
Several mechanisms were proposed early on,   in particular mechanisms based on 
 spin fluctuations\cite{Emery,ScalapinoSpin, Emery2, Pines},  
 charge fluctuations\cite{Ruvalds,Varma,Hirsch},   and other more exotic ideas such 
 as resonating valence bonds\cite{Anderson}.    
 A more recent work studies the possibility of superconductivity at small coupling,
 where here the mechanism goes back to ideas of Kohn-Luttinger\cite{Raghu,Kohn}.  
 
Since the Mott-insulating anti-ferromagnetic phase at half-filling is well understood,
much of  the theoretical literature attempts  to understand  understand  how doping
``melts''  the anti-ferromagnetic  (AF) order,  and how the resulting state can become 
superconducting.    This has proven to be difficult to study,  perhaps in part due to
the fact that AF order is spatial,  whereas superconducting order is in
momentum space.     (For a review and other refereces,  see \cite{Wen}.)    
Furthermore,  superconductivity exists at  reasonably  low densities far from the AF order at half-filling,  which
suggests that 
that one can perhaps  treat the model as a gas,  with 
superconductivity arising as a condensation of Cooper pairs  as in the BCS theory, 
and we will adopt this point of view in the present work.

\begin{figure}[htb] 
\begin{center}
\hspace{-15mm} 
\includegraphics[width=7cm]{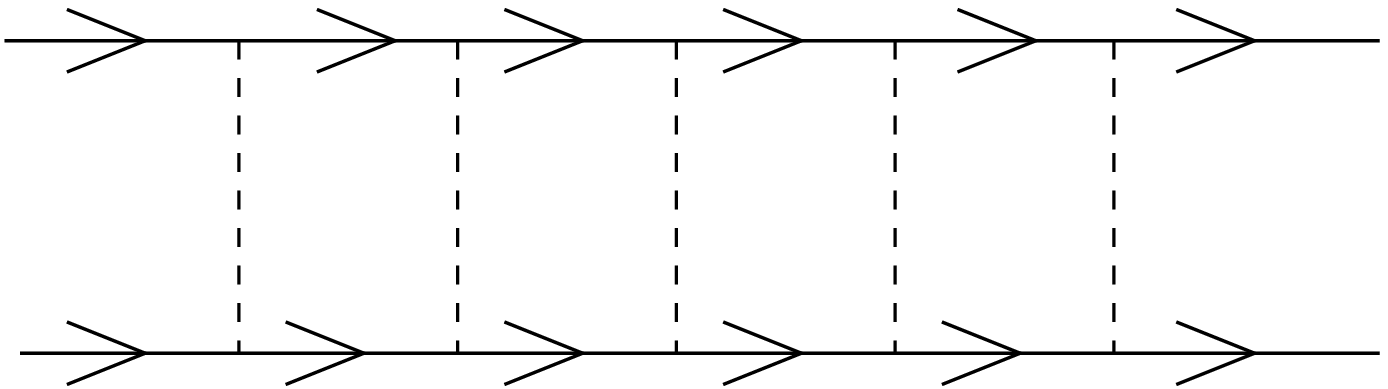} 
\end{center}
\caption{Feynman diagrams that lead to the BCS pairing instability,  where dashed lines are 
phonons.}  
\vspace{-2mm}
\label{ladders} 
\end{figure}

Since there is as yet no consensus on the precise pairing mechanism in the cuprates, 
it is worthwhile continuing the search for new ones.   In this paper we 
take a very conservative approach within the Hubbard model,  wherein we 
do not postulate any particular quasi-particle excitations that would play the role of
the phonons,  and we ignore the AF order at very  low doping.   
In a field-theoretic approach to the BCS theory,   the pairing instability arises from the sum
of the ladder diagrams for phonon exchange,  as shown in Figure \ref{ladders}.    The 
sum of these diagrams has a pole at energy equal to the gap,  and this pole is sufficient to cause
the instability,   and signifies  gapped  charge 2 excitations.
(See e.g. \cite{BCS,MahanBook}).  
This will be reviewed in section III.

\begin{figure}[htb] 
\begin{center}
\hspace{-15mm} 
\psfrag{mkd}{$-\kvec\down$}
\psfrag{ku}{$\kvec\up$}
\psfrag{mkpd}{$-\kvec'\down$}
\psfrag{kpu}{$\kvec'\up$}
\includegraphics[width=10cm]{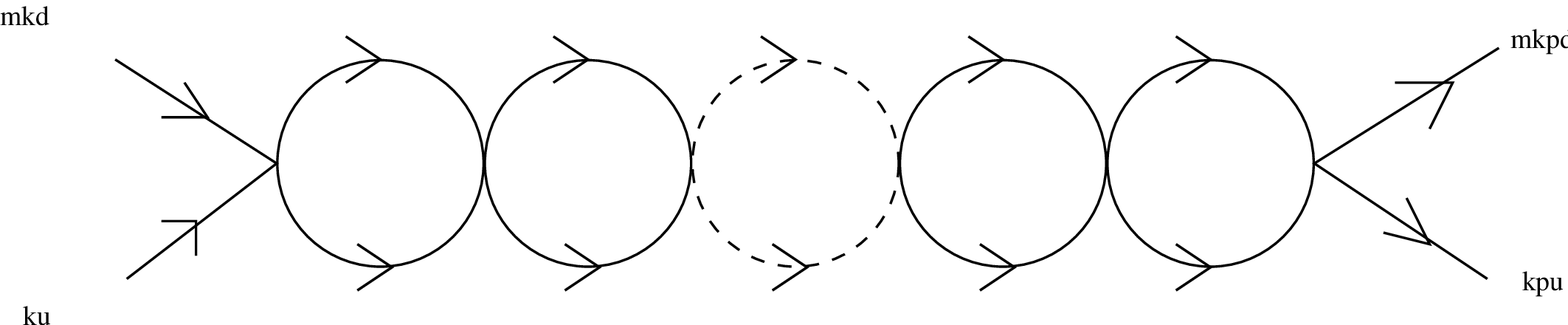} 
\end{center}
\caption{Feynman diagrams contributing to $\Veffscreen$.}  
\vspace{-2mm}
\label{Vscreen} 
\end{figure}

\begin{figure}[htb] 
\begin{center}
\hspace{-15mm} 
\psfrag{mkd}{$-\kvec\down$}
\psfrag{ku}{$\kvec\up$}
\psfrag{kpd}{$\kvec'\down$}
\psfrag{mkpu}{$-\kvec'\up$}
\includegraphics[width=10cm]{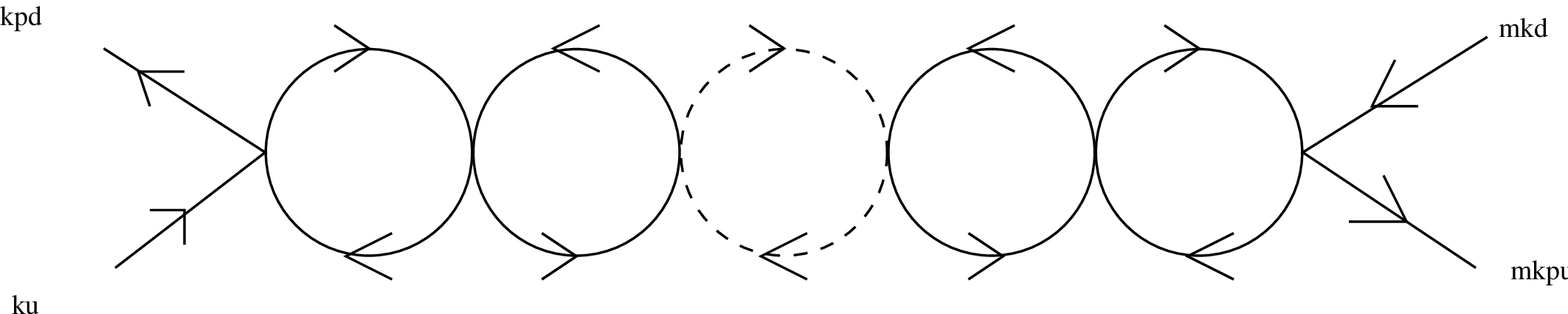} 
\end{center}
\caption{Feynman diagrams contributing to $\Veffex$.}  
\vspace{-2mm}
\label{Vex} 
\end{figure}

  Suppose that the electron-phonon interaction  
were treated as an effective,  short-ranged electron-electron interaction,  so that  exchange of
a single phonon were replaced by an effective 4-electron vertex.    Then the diagrams in 
Figure \ref{ladders}  become the diagrams in Figure \ref{Vscreen}.     In this work we explore
the possibility of pairing instabilities arising from certain similar  classes of Feynman diagrams in the Hubbard
model.    
Whereas the diagrams in Figure \ref{Vscreen}  lead to the BCS pairing instability for
bare {\it attractive} interactions (see below),   for repulsive interactions they merely screen
the strength of the Coulomb interaction.   However,  as we will show,   the sum of another class of diagrams
shown in Figure \ref{Vex}  do indeed exhibit poles  possibly signifying pairing instabilities very near
the Fermi surface.  An obvious criticism of the present work is that for the cuprates,  the coupling is
large,  and one should not trust perturbation theory.  In answer to this,  the diagrams we will focus upon
can be summed up, and have a well-defined limit as the coupling goes to infinity.   Furthermore,
since we are focussing on sums of diagrams that lead to poles in the effective pairing potential,   
it is possible that these singular diagrams dominate the perturbative expansion.     
Certainly our calculation can be improved upon;   we wish here to present a possible pairing
mechanism in its simplest form.     For instance,  we ignore the effect of interactions on the 
quasi-particle energies,  and take $\xi_\kvec$ to be that of the free theory,  ignoring self-energy
corrections,  which are known to be significant in the pseudo-gap region.    One can repeat the
analysis  for instance by extracting $\xi_\kvec$ from experiments,  however we leave this for the future.  
At this stage,  it is more useful to study the mechanism and its plausibility in its simplest form.   

Comments on the  connection between the present work and our earlier one\cite{HubbardGap}  are called for.
In \cite{HubbardGap}  the diagrams in Figure  \ref{Vscreen}  at {\it zero chemical potential and temperature}  were
viewed as contributions to an effective interaction between 2 electrons in vacuum,  and it was shown that 
there are regions in the Brillouin zone where this interaction is attractive.    This effective interaction 
was then fed into a BCS gap equation,  which showed solutions.     Our current understanding is that
this may not be consistent for the following reasons.    Superconductivity is a phenomenon that relies
strongly on properties near the Fermi surface,  Fermi blocking, etc,   and thus effectively attractive interactions
in free space,  though they may lead to bound states,  are believed to be insufficient to cause superconductivity.   
In this paper the diagrams in Figures \ref{Vscreen}, \ref{Vex}  are calculated at finite temperature and chemical potential,
and can thus be studied near the Fermi surface.   In fact,  the diagrams in Figure \ref{Vex}  are zero at zero density.   
Although in a different channel than the diagrams in Figure \ref{Vscreen},  they still contribute to Cooper-pair 
Green functions,  and poles in these functions could signify pairing just as the poles in the diagrams in the 
direct channel Figure \ref{Vscreen}.    
 In summary,  as far as superconductivity is concerned,   there is no connection
between the present work and \cite{HubbardGap}.       Furthermore,  it was suggested that the results in \cite{HubbardGap}
were perhaps more relevant to the so-called pseudo-gap,  since the solutions of the gap equation increased
all the way down to zero doping.

The remainder of this paper is organized as follows.   In the next section we define the
Cooper `pairing'  Green functions of interest.  
In section III we first review how the Cooper pairing instability manifests itself as poles  in such
Green's functions in the BCS theory.    The remainder of this section specializes to the 
Hubbard model,   where analogous poles are found for a different class of diagrams.   
The latter diagrams have been studied before in the  context of  Landau damping 
and plasmons.    
However,  in the present work these diagrams are used to define an effective potential,  
which leads to a different  dependence on kinematic variables near the Fermi surface.
Thus,  the detailed effective potential studied in this paper has not been considered before 
in connection with pairing;    this is explained in  detail
below.    
The gap is related to the location of these poles,  and solutions are found numerically. 
For $U/t = 10$,  we find non-zero gap solutions in the anti-nodal directions 
 in the range of hole doping  $0.03<h<0.24$ with a dome-like shape. The maximum of the gap 
 is approximately $\Delta/t  = 0.08$   and  occurs around $h=0.11$.

\section{Green functions and the effective  Cooper pair potential}

We study the two-dimensional Hubbard model:
\beq
\label{HubbardHamiltonian}
H = -t \sum_{ <i,j>, \alpha = \up, \down}  c^\dagger_{i,\alpha}  c_{j, \alpha}  
+U \sum_i  n_{i, \up}  n_{i, \down}  
\eeq
where $i,j$ label  sites of a two-dimensional square lattice,  $<i,j>$  denotes nearest  neighbors,  and 
$n_{i, \alpha} = c^\dagger_{i, \alpha} c_{i, \alpha}$.    Our convention for neighboring interactions is 
e.g. $\sum_{<i,j>}   c^\dagger_i  c_j  =  c^\dagger_1 c_2  + c^\dagger_2 c_1 + .....$ such that $H$ is hermitian.   

The lattice operators $c_{\rvec_i, \alpha}$ where $\rvec_{i}$ is a lattice site,  will correspond to the continuum
fields
\beq
\label{fields}
\psi_\alpha (\xvec, t )  =  \int  \frac{ d^2 \kvec}{2\pi}  \,  c_{\kvec, \alpha}  (t)   \,  e^{i \kvec \cdot \xvec }
\eeq
The free hopping terms are diagonal in momentum space: 
\beq
\label{Hfree}
H_{\rm free}  =  \int d^2 \kvec \,  \omega_\kvec  \sum_{\alpha = \up, \down}   c^\dagger_{\kvec, \alpha}   c_{\kvec, \alpha}   
\eeq
with  the 1-particle energies
\beq
\label{omegak}
\omega_\kvec =  - 2 t (\cos (k_x a )  + \cos (k_y a)  )  \eeq
where $a$ is the lattice spacing.   
We henceforth scale out the dependence on $t$ and $a$ such that all energies,  temperatures and 
chemical potentials  are in units of $t$.    
The interactions then depend on the dimensionless coupling  $g \equiv  U/t$,  which is positive for
repulsive interactions.       
Henceforth $\xi_\kvec \equiv \omega_\kvec - \mu$ is the 1-particle energy measured relative to the
Fermi surface,  where $\mu$ is the chemical potential.    

In order to probe possible Cooper pairing instabilities,  we consider the Green function
$\langle \psi^\dagger_\up (x_1)  \psi^\dagger_\down (x_2)  \psi_\down (x_3)  \psi_\up (x_4 ) \rangle 
$
where $x = (\xvec, t)$.   
The Fourier transform in both space and time of these  functions are correlation functions of
the operators $c^\dagger_{\kvec,\alpha} ( { \smll \xi }) =  \int  \frac{dt}{\sqrt{2\pi}}   e^{-i \xi t }  c^\dagger_{\kvec, \alpha} (t) $,
and their hermitian conjugates.    We  study  Green functions specialized to Cooper pairs:
$
\langle  c^\dagger_{-\kvec' \up} ({ \scriptstyle \xihat '})  c^\dagger_{\kvec' \down}  ( {\scriptstyle \xi' })  c_{-\kvec \down}  ( {\scriptstyle \xihat })  
c_{\kvec \up}  ( {\scriptstyle \xi} ) \rangle 
$.
It should be emphasized that since the above Green functions are just Fourier transforms of the spatial/temporal Green functions,  although the $\xi$ are energy variables, 
 they are not necessarily 
`on-shell',  i.e. $\xi$ is not necessarily $\xi_\kvec$  (borrowing the relativistic terminology).     The only constraint is energy conservation:
$\xihat + \xi = \xihat'  + \xi'$.    

There are two important types of quantum corrections to the vertex,  which we denote as
$  \Veffscreen $ and $ \Veffex $,  where,  for reasons explained below, 
  {\it sc} refers to {\it screened} and {\it ex}  to
{\it exchange}.     $\Veffscreen$ is defined as 
\beq
\label{Vscreendef}
\Veffscreen  =  
\langle  c^\dagger_{-\kvec' \up} ( {\smll \xihat'} )  c^\dagger_{\kvec'\down}  ( {\smll \xi'} )   c_{-\kvec \down}  ({ \smll \xihat } )  
c_{\kvec \up}  ( {\smll \xi }  ) \rangle_{\rm trunc}^{(sc)}  
\eeq
where {\it trunc}  refers to the truncated Green function, i.e.  stripped of external propagators and
energy-momentum conserving delta functions,  and {\it sc}  refers to the diagrams in Figure \ref{Vscreen}.  
Incoming (outgoing) arrows correspond to annihilation (creation) operator fields.   
The other class of diagrams are shown in Figure \ref{Vex} and define $\Veffex$  
as in eqn.  (\ref{Vscreendef}).

When `on-shell',  which is to say that frequencies are  1-particle energies, i.e. $\xi = \hat{\xi} = \xi_\kvec$ and $   \xi' = \hat{\xi}' = \xi_{\kvec'}$,
then these truncated Green functions contribute  to the matrix element,  i.e. form-factor,  of the 
integrated interaction hamiltonian density:
\beq
\label{VonShell}
V(\kvec, \kvec' ) =  \int d^2 \xvec \langle -\kvec'\up , \kvec'\down | \CH_{\rm int} (\xvec ) |  \kvec\up, -\kvec\down \rangle
\eeq
Since there is no integration over time in the above equation,   $\xi_\kvec$ and $\xi_{\kvec'}$ are not necessarily 
equal.     To lowest order,  $V = g$.    The  form factor $V^{\rm (sc) } (\kvec, \kvec')$   corresponding to  the diagrams of 
$\Veffscreen$  is
equal to $\Veffscreen$ (as computed below)  with $\xi \to (\xi_\kvec + \xi_{\kvec'} ) /2$\cite{HubbardGap},   whereas $V^{\rm (ex)}$
is simply $\Vex$ placed on shell.    For both,  one has the necessary symmetry:
$V (\kvec, \kvec'  )  =  V (\kvec' , \kvec )$.          

The evaluation of $\CV_{\rm eff}^{(sc, ex)}$ at finite density and temperature is standard,  however  for 
completeness we provide some details.   
Consider first $\Veffscreen$.    These diagrams factorize into 1-loop integrals and form a geometric series.  
There is no fermionic minus sign coming from each loop  since the arrows do not form a
{\it closed} loop.     Momentum conservation at each vertex gives a loop integral that is independent of
$\kvec, \kvec'$:
\beq
\label{Vscsum}
\Veffscreen (\xi  )  =  \frac{  g}{1- g \Lsc  (\xi  )}
\eeq
where 
$\Lsc$ is the one-loop integral:
\beq
\label{Lscdef}
\Lsc (\xi )  =   - T \sum_n \int  \frac{d^2 \pvec}{(2\pi)^2}  \(  \inv{ i \nu_n - \xi_\pvec }   \)
\( \inv{ 2 \xi    - i \nu_n - \xi_{-\pvec}  } \)
\eeq
where $T$ is temperature,  $\nu_n$ is a fermionic Matsubara frequency,  $\nu_n = 2\pi (n+\inv{2}) T$ with
$n$ an integer,  and $\xi_\pvec =  \omega_\pvec - \mu$ where $\mu$ is the chemical potential.   
One needs the following identity:
\beq
\label{scident}
T \sum_n  \( \inv{i\nu_n - \xi_\pvec} \) \(   \inv{  2 \xi  - i\nu_n - \xi_{-\pvec} }\) 
= \frac{   f(\xi_\pvec) -1/2 }{\xi    -   \xi_\pvec}
\eeq
where  $f(\xi)  =  1/ (e^{\xi/T} +1) $  is the fermionic occupation number,  and  we have used $\xi_\pvec = \xi_{-\pvec}$.   The above identity is 
valid before analytic continuation from imaginary to real time,  i.e. when $2 \xi  $  is twice a fermionic Matsubara frequency. 
The final result is then:
\beq
\label{LscFinal}
\Lsc (\xi )  =   \inv{2}  \int   \frac{d^2 \pvec}{(2\pi)^2} 
\(  \frac{   1  } { \xi    -   \xi_\pvec  +i \eta } \)  \tanh(\xi_\pvec /2T)
\eeq
where $\eta$ is infinitesimally small and positive.   
Integration is over the first Brillouin zone,  $-\pi  \leq  p_{x,y}  \leq \pi$.

The diagrams for $\Veffex$,  though simply an `exchanged' version of those for $\Veffscreen$,  have
a rather different  and more complicated structure.    Here there is a fermionic minus associated with each loop since
the arrows form a closed  loop.   Momentum and energy conservation at each vertex now leads
to a loop integral that depends on $\qvec \equiv \kvec' - \kvec$ and $\Deltaxi \equiv  \xi' - \xi$:
\beq
\label{Vexddef}
\Veffex (\qvec, \Deltaxi ) =   \frac{g}{1- g \Lex}
\eeq
where 
\beq
\label{LexdefT}
\Lex (\qvec,  \Deltaxi )  =   T \sum_n  \int  \dpvec 
\( \inv{ i \nu_n - \xi_\pvec } \)  \(   \inv{  \Deltaxi + i \nu_n  -  \xi_{\pvec + \qvec}}\) 
\eeq
Now one needs the identity
\beq
\label{ident2}
T \sum_{n} \(  \inv{i \nu_n - \xi_\pvec} \) \( \inv{i\nu_n + i \omega_m - \xi_{\pvec+ \qvec} } \) = 
\frac{ f(\xi_\pvec) - f(\xi_{\pvec+ \qvec} ) }{i \omega_m + \xi_\pvec - \xi_{\pvec+ \qvec} }
\eeq
which is valid for $\omega_m $ twice a bosonic Matsubara frequency.  
After analytic continuation $i\omega_m \to   \Deltaxi + i \eta $,  one has
\beq
\label{Lexdef}
\Lex (\qvec,  \Deltaxi )  =  \int   \dpvec  
\(  \frac{ f(\xi_\pvec )  - f(\xi_{\pvec  + \qvec} ) }{\Deltaxi + \xi_\pvec - \xi_{\pvec + \qvec}  +i \eta } \)  
\eeq

In the above formulas it  is implicit  that  $\CV^{(\rm sc, ex)}$ is defined by the real part of $L^{\rm (sc, ex)}$ 
in the limit  $\eta \to 0$.      
In  the limit of zero density, i.e.  $f=0$,  note that $\Vex =0$,  whereas  $\Veffscreen$ is non-zero.

\section{Cooper pairing instabilities as poles in the effective potential}.

It is well-known that the Cooper pairing instability of the BCS theory can be
understood as a pole in $\Veffscreen$\cite{BCS,MahanBook}.     This can be seen by inserting a complete set of states between
the pair creation and annihilation operators in eqn. (\ref{Vscreendef}),  and noting that
poles in the energy $\xi$ signify  bound states of electric charge $2$,  which are interpreted as the Cooper pairs.  
To see that this gives the correct result for the gap,  let us leave behind the Hubbard model and  specialize the coupling $g$
to that of the BCS theory,  where it arises from the interaction with phonons.   For simplicity, 
$g = - |g|$ is taken to be negative,  signifying attractive interactions,  in a narrow region near the 
Fermi surface   $| \xi | < \omega_D$,  where $\omega_D$ is the Debye frequency,  otherwise zero.  
We approximate the integral over $\pvec$ near the Fermi surface as   $\int  \rho (\xi_\pvec )  d \xi_\pvec 
\approx  \rho (0) \int d \xi_\pvec$,   where $\rho (\xi )$ is the density of states.    Letting $\xi=\Delta/2$, 
in the limit of zero temperature, 
\beq
\label{LscBCS}
\Lsc \approx    \frac{\rho (0)}{2}  \( 
\int_{0<\xi_\pvec < \omega_D}    d\xi_\pvec  \inv{\Delta/2 - \xi_\pvec}    - 
\int_{-\omega_D <\xi_\pvec < 0 }    d\xi_\pvec  \inv{\Delta/2 - \xi_\pvec}  \) 
\approx \rho(0) \log \( \frac{\Delta}{2\omega_D}  \) 
\eeq
(The first integral should be interpreted as the principal value.)
There is a  pole in $\Veffscreen$  because  both  $\Lsc$ and $g$ are negative.   The location of the pole is 
given by  $1 + |g| \rho (0) \log (\Delta/2 \omega_D) =0$,
which leads to the standard result for the gap:   
$\Delta =  2 \omega_D  \exp (-1/ ( |g| \rho(0)) )$.   Note that there is a gap $\Delta$ for arbitrarily small  negative coupling  $g$.

\def\LexOnShell{\hat{L}^{(ex)}}

Returning to the  positive $g$  Hubbard model,  i.e. with repulsive interactions,  there are  no poles in $\Veffscreen$ since $\Lscreen$  turns out to be  negative at finite temperature and density.      These  quantum corrections  merely screen the Coulomb potential.  It is meaningful to define 
a screened coupling near the Fermi surface as  $\gR = \Veffscreen(\xi= 0)$.  
We wish to point out that at zero density,  i.e. $f=0$,  and $T=0$,   $\Lscreen$ can become positive,
which means that $\Veffscreen$ can be attractive.   This was explored in \cite{HubbardGap},  
and with the help of a hypothesized gap equation,  one finds anisotropic solutions with properties
suggestive of the pseudo-gap.   

We now  turn to the investigation of the possibility of analogous  pairing instabilities  in  $\Veffex$.   
By construction,  since it contributes to the full effective potential in the same way $\Veffscreen$ does,
then for the same reason as above,  poles in $\Veffex$ could signify Cooper pairs.  
 Poles in $\Veffex$ for positive $g$  only exist if $\Lex$ can
be positive.        For $\Deltaxi =0$,   the integrand  in $\Lex$ is always negative,  thus $\Lex$ is negative and there
are no poles for positive $g$.    However we are precisely interested in the case $\Deltaxi \neq 0$ since
$\Deltaxi $ represents a difference of incoming and outgoing energies,  and  it is thus poles 
at non-zero  values of $\Deltaxi$ that could signify Cooper paired bound states.   As we will see,
there are indeed regions of the parameter space near the Fermi surface where $\Lex$ is positive.      
For $\qvec$ and $\Deltaxi$  unrelated,   the integral is easily evaluated numerically,  and has a smooth, 
well-defined limit as $\eta \to 0$.     However physically we are interested in this
function {\it  on-shell},  i.e. the corresponding form factor $V^{\rm (ex)}$    with $\Deltaxi = \xi_{\kvec'} - \xi_\kvec $ when $\qvec = \kvec' - \kvec$,  so that $\Veffex$ now depends on $\kvec, \kvec'$,  not only $\qvec$.    With this identification 
of $\Deltaxi$,   $V^{\rm (ex)}$ is properly viewed as an effective potential,   as is  $\Veffscreen$.   
For clarify,  let us define the on-shell version of $\Lex$ that plays an important role in this paper as follows:
\beq
\label{Lon}
\LexOnShell ( \kvec,  \kvec' )  =   \Lex(  \qvec=\kvec' - \kvec, \, \Deltaxi = \xi_{\kvec'} - \xi_\kvec) 
\eeq

Versions of the  function $\Lex (\qvec, \Delta_\xi)$  appear in  other essentially different physical contexts, 
and we wish to clarify this point.   The analysis in  \cite{ScalapinoSpin,Emery2}  involves the  same diagrams that define
$\Lex$,   however with the important difference that there $\Delta_\xi = 0$;   this is here interpreted as the effective potential 
$\Veffex$ with both $\kvec, \kvec'$ on the Fermi surface and  is thus  a different function of
$\kvec, \kvec'$ than  
our on-shell $\LexOnShell$.  
We also point out that the function $\Lex$ off-shell,  i.e. with $\Delta_\xi $ equal to
an arbitrary frequency $\omega$ unrelated to $\qvec$  appears in the RPA expression for
the dielectric response function   $\vep_{\rm RPA}  (\omega, \qvec ) =  1-  g \Lex ( \Delta_\xi = \omega,  \qvec )$\cite{MahanBook}.     Here,  $\omega$ is the frequency of an external probe,  such as an electric field.   Solutions  $\omega (\qvec)$  to the
equation 
$\vep_{\rm RPA}  (\omega (\qvec) , \qvec)= 0$ are interpreted as Landau damping.   
Plasmons,  i.e. quantized electric charge fluctuations,  are manifested as 
delta-function peaks in $- \Im (1/\vep_{RPA} )$  as a function of $\omega$.   
The pairing mechanism studied here  thus  appears  closest to the idea of plasmon mediated 
superconductivity\cite{Ruvalds,Mahan}.    However our analysis differs in important ways:  no  low energy plasmons were postulated,  and once again,  the properties  of the on-shell $1-g \LexOnShell$ are very different from those of  
$\vep_{\rm RPA}$.

The integral defining $\LexOnShell$ is rather delicate in comparison with $\Lsc$ and the off-shell $\Lex$.     
In this case there is a pole in the integrand when $\pvec = \kvec$.   Thus the $\pvec$ integral 
should be understood as the Cauchy principal value (PV).    Namely,  inside an integral over a variable
$x$,  one has the identity:
\beq
\label{CPV}
\lim_{\eta \to 0^+}  \inv{x + i \eta  }   = {\rm  PV}   \( \inv{x}  \)   - i \pi \delta(x) 
\eeq
Recall the PV is defined by excising a small region around the pole:
$\int d p_x  \to  \lim_{\delta \to 0} \( \int_{-\pi}^{k_x - \delta} dp_x  +  \int_{k_x + \delta}^\pi  dp_x \)$
and similarly for  $\int dp_y$.

In order to better understand other features of the function $\Lex$ on-shell,     
it  is instructive to study the one-dimensional version of $\Lex$,  appropriate to 
a chain of lattice sites,       since here  the integral can be performed 
analytically.   This is described in the Appendix.     (It should be emphasized that this
exercise is not meant to capture the physics of the one-dimensional Hubbard model,
which is exactly solvable\cite{Korepin},  but rather simply to gain intuition on the function $\Lex$ in
two dimensions.) 
As  shown analytically  in the appendix, 
the on-shell $\Lex$ actually diverges as the temperature goes to zero in one dimension.     
The reason is that in the limit of zero $T$,  the occupation number $f$ is a step function,
and the $\int_{k_F + \delta}^\pi  d\pvec$ piece of the PV is absent,  leading it to be ill-defined.   
The temperature $T$ should thus be viewed as an infra-red regulator in one dimension.
 In two dimensions we will also not take $T\to 0$ for analogous reasons.     For arbitrarily small $T$,  we verified that $\Lex$ is well defined numerically as 
$\delta \to 0$.      As shown in the appendix,   one can demonstrate analytically that 
 there is a narrow region around the Fermi surface where 
$\Lex$  is positive and thus $\Veffex$  has poles.    As we now show,   the same is true in two dimensions.

\begin{figure}[htb] 
\begin{center}
\hspace{-15mm} 
\psfrag{mkd}{$-\kvec\down$}
\psfrag{kx}{$k_x$}
\psfrag{ky}{$k_y$}
\psfrag{q}{$\qvec$}
\psfrag{kp}{$\kvec'$}
\psfrag{kF}{$\kvec_F$}
\psfrag{FS}{${\rm \scriptstyle{Fermi ~surface}}$}
\psfrag{th}{$\theta$}
\includegraphics[width=8cm]{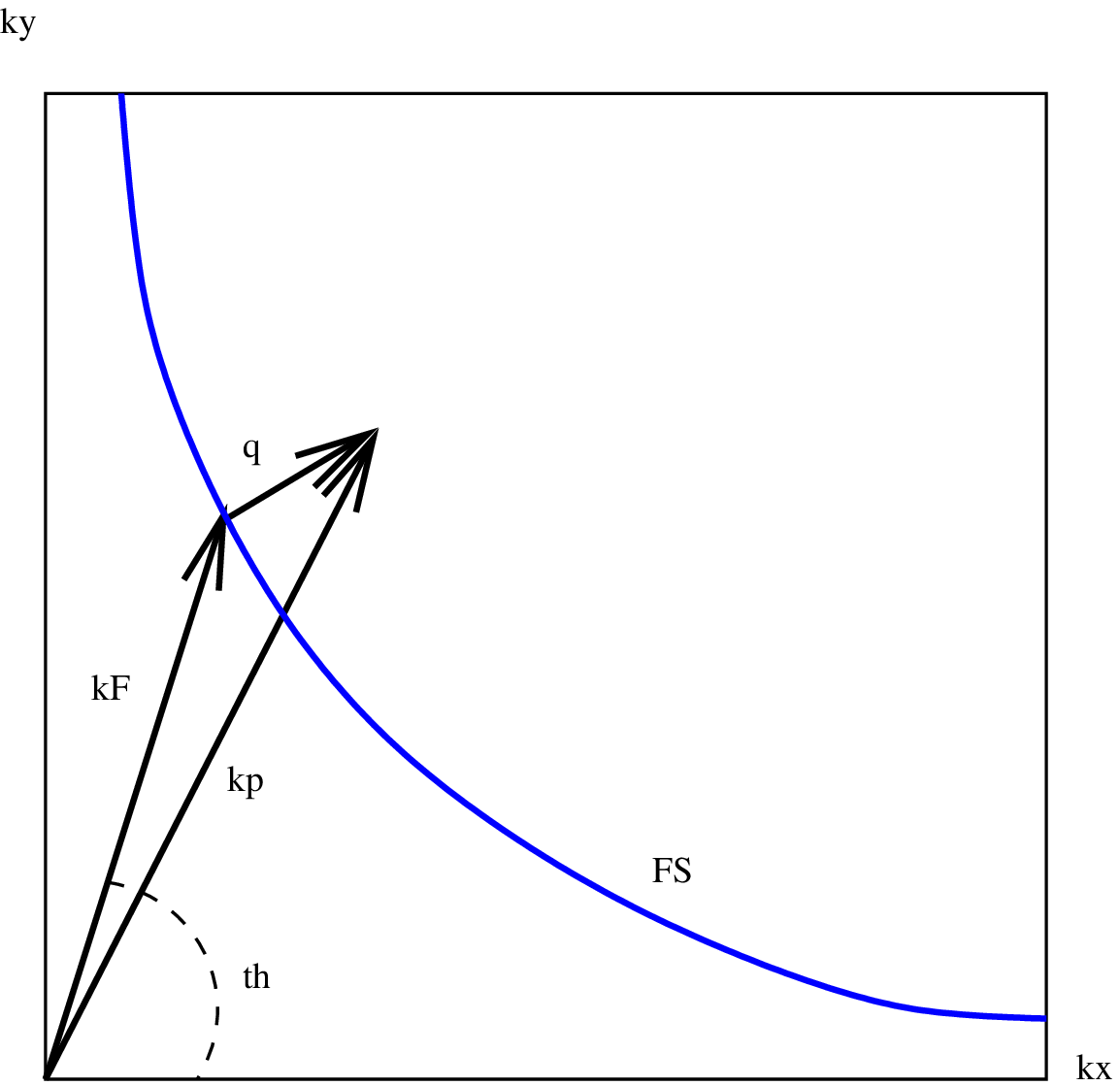} 
\end{center}
\caption{Geometry near the Fermi surface in the first quadrant of the Brillouin zone.}  
\vspace{-2mm}
\label{FS} 
\end{figure}

We are interested in potential instabilities when $\kvec, \kvec'$ are near the Fermi surface.  
To be more specific,  let $\kvec = \kvec_F (\mu)$ be right on the Fermi surface,  i.e. $\xi_\kvec=0$,
and $\kvec'$ be slightly off the surface,  with some small $\xi_{\kvec'}$,   so that now  $\Deltaxi = \xi_{\kvec'}$.  
See Figure \ref{FS}.   
As usual,  the $\theta = \pi/4$ direction will be referred to as nodal,  whereas the
$\theta = 0$ as anti-nodal.    
For a fixed value of the chemical potential,  $\LexOnShell (\kvec_F, \kvec' )$ depends only
on  $|\kvec'|$ and  the angular directions $\theta, \theta'$ of $\kvec_F, \kvec'$.   The magnitude $|\kvec'|$   
can be related to $\Deltaxi $ as follows:  $\Deltaxi = \omega_{\kvec'} - \mu$.      At fixed chemical potential,  with $\kvec$ on the Fermi surface, 
we can  thus view $\LexOnShell$ 
as a function  only of $\theta, \theta'$ and $\Deltaxi$.

The chemical potential will be  related to the hole doping $h$ by the formula:
\beq
\label{hdef}
1- h =  2 \int \frac{d^2 \kvec} {(2\pi)^2 } \,  \inv{  e^{\xi(\kvec)/T} +1 }
\eeq
This is clearly an approximation since there are self-energy corrections that modify $\xi_\kvec$ which
we ignore;    however,  since the experimentally measured Fermi surfaces can be fit to 
a $\xi_\kvec$ of a tight-binding form,  we do not expect these corrections to drastically affect 
the main features of our results. 
Henceforth,  we express various properties in terms of $h$ defined by the above
equation in the limit $T\to 0$.

\begin{figure}[htb] 
\begin{center}
\hspace{-15mm} 
\psfrag{x}{$\Deltaxi$}
\psfrag{y}{$\LexOnShell$}
\includegraphics[width=7cm]{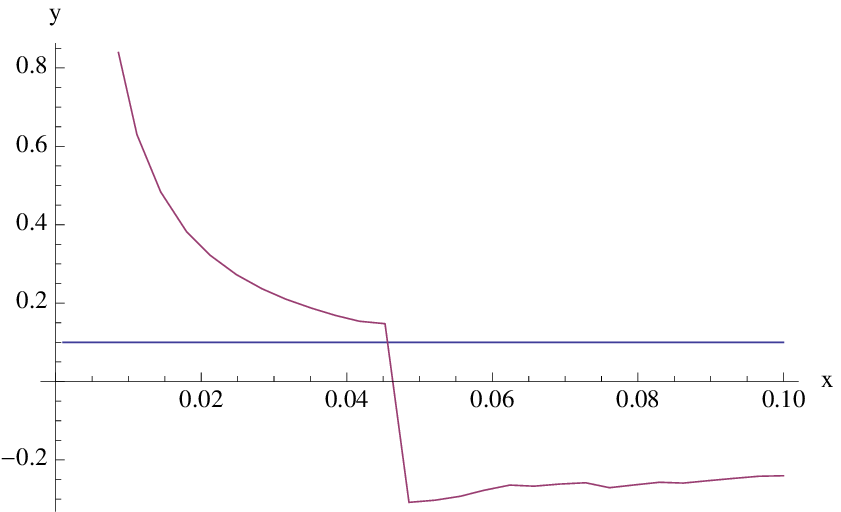} 
\end{center}
\caption{The loop integral $\LexOnShell$  in the anti-nodal direction as a function of $\Deltaxi$ for hole doping $h=0.15$ and temperature $T_0=0.001$.
The straight line represents $1/g =0.1$ and where it intersects $\LexOnShell$ gives the value of the gap $\Deltagap$.}  
\vspace{-2mm}
\label{Lexh2} 
\end{figure}

Let us first illustrate our findings at the fixed doping $h=0.15$.  
In Figure \ref{Lexh2}  we plot $\LexOnShell$ for both $\kvec,  \kvec'$ in the anti-nodal direction as 
a function of $\Deltaxi$ 
at the low temperature  $T_0=0.001$  (compared to the bandwidth).   
One sees that for small enough $\Deltaxi$,  i.e. close enough to the Fermi surface,  
$\LexOnShell$  becomes positive.    As  in the case of the BCS instability reviewed above,   let  us identify
the gap $\Deltagap=\Deltaxi$ with the location of the pole in $\Veffex$ in the anti-nodal direction,  i.e. the solution to the equation:
\beq
\label{gapeqn}
\inv{g}  =   \LexOnShell ( \Deltagap, \mu,  T)
\eeq
where it is implicit that $\theta = \theta' =0$.   
For $\gsc = 10$,   appropriate to the cuprates, 
one sees from Figure \ref{Lexh2} that $\Deltagap \approx 0.045$. 
   
Two other features are apparent from Figure \ref{Lexh2}:  

\medskip
\noindent
(i)  There is a solution to
eqn. (\ref{gapeqn}) for arbitrarily large $\gsc$ since $\LexOnShell$ passes through zero.  
Furthermore,  as $\gsc \to \infty$,  $\Deltagap $ saturates to where $\LexOnShell$ crosses the real axis,
in this case $\Deltagap \approx  .047$.   This is reminiscent of claims of superconductivity 
in the t-J model\cite{Maier},  since it is a strong coupling version of the Hubbard model, 
and  since the interactions considered here are also  instantaneous. 
\medskip

\noindent
(ii)  Since $\LexOnShell$ has a maximum,  there are only solutions to eqn. (\ref{gapeqn}) for
$\gsc$ larger than a minimum value,  in this case $\gsc$  greater than approximately  $0.2$. 
Since no such threshold was found in \cite{Raghu},   one should conclude that the pairing 
mechanism presented here is essentially different.   

The solutions  $\Deltagap$  to eqn.  (\ref{gapeqn})  depend on the temperature.  However
since we  view  $T$ as an infra-red regulator,  one should think of this in terms of 
the renormalization group.   Namely,   $g$ can be made to depend on $T$ in 
such a manner as to keep the solution $\Deltagap $ fixed.   
One finds that $g$ increases with decreasing $T$.   For instance,  at  $h=0.15$, 
if $\gsc = 1$ at $T_0 = 10^{-3}$,   then $\gsc  \approx 2.3$ for $T_0 = 10^{-4}$.  
We emphasize that there is only one free parameter in our calculation,  the value of $g$ at the reference temperature $T_0$.  
Henceforth we fix the reference  temperature  $T_0 = 0.001$,  keeping $\gsc =10$.

\begin{figure}[htb] 
\begin{center}
\hspace{-15mm} 
\psfrag{x}{$\Deltaxi$}
\psfrag{y}{$\LexOnShell$}
\psfrag{h1}{$h=0.15$}
\psfrag{h2}{$h=0.24$}
\includegraphics[width=10cm]{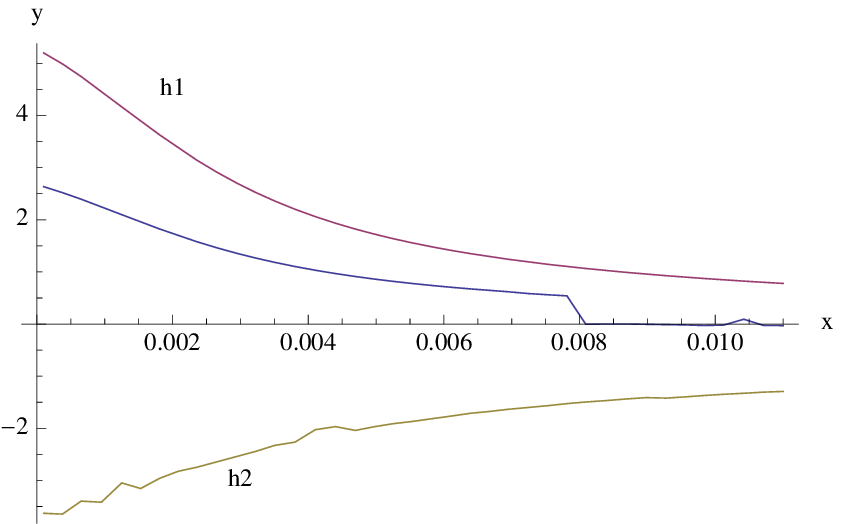} 
\end{center}
\caption{The loop integral $\LexOnShell$  in the anti-nodal direction as a function of $\Deltaxi$ for hole doping $h=0.05, 0.15$ and $0.24$.}  
\vspace{-2mm}
\label{Lexh3} 
\end{figure}

\begin{figure}[htb] 
\begin{center}
\hspace{-15mm} 
\psfrag{y}{$\Deltagap$}
\psfrag{x}{$h={\rm hole ~ doping}$}
\includegraphics[width=7cm]{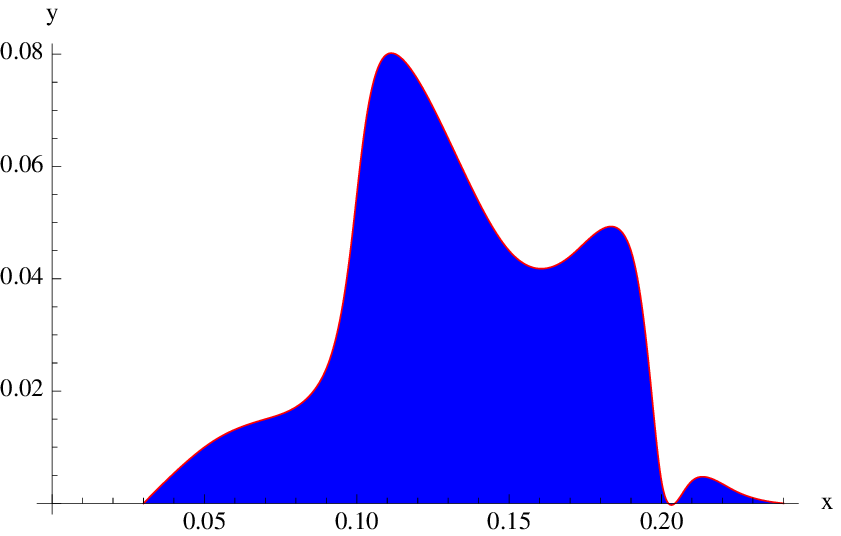} 
\end{center}
\caption{$\Deltagap$ in the anti-nodal direction  as a function of hole doping $h$.}  
\vspace{-2mm}
\label{Gapofh} 
\end{figure}

In order to understand how $\Deltagap$ depends on doping,  in Figure \ref{Lexh3} we plot
$\LexOnShell$ for $h=0.05, 0.15$ and $0.24$.    One sees that as $h$ is decreased,  
$\LexOnShell$ crosses zero at smaller values of $\Deltaxi$,  signifying a smaller $\Deltagap$.   
Finally,  when $h$ is large enough, $\LexOnShell$ is nowhere positive,  signifying no solution to
eqn.  (\ref{gapeqn}).      In Figure \ref{Gapofh}  is shown the anti-nodal  solution $\Deltagap$ as a function 
of doping $h$.    The largest gap  $\Deltagap =0.08$  occurs around $h=0.11$.
For the cuprates,  $t \approx 0.4\, {\rm eV}$,  which gives $\Deltagap = 32\, {\rm meV}$,  which 
compares favorably with experiments.  
  It is important to note that reasonable values for $\Deltagap$ were obtained
without introducing any explicit cut-off in momentum space related to the bandwidth.

\begin{figure}[htb] 
\begin{center}
\hspace{-15mm} 
\psfrag{y}{$\Deltagap$}
\psfrag{x}{$g$}
\includegraphics[width=7cm]{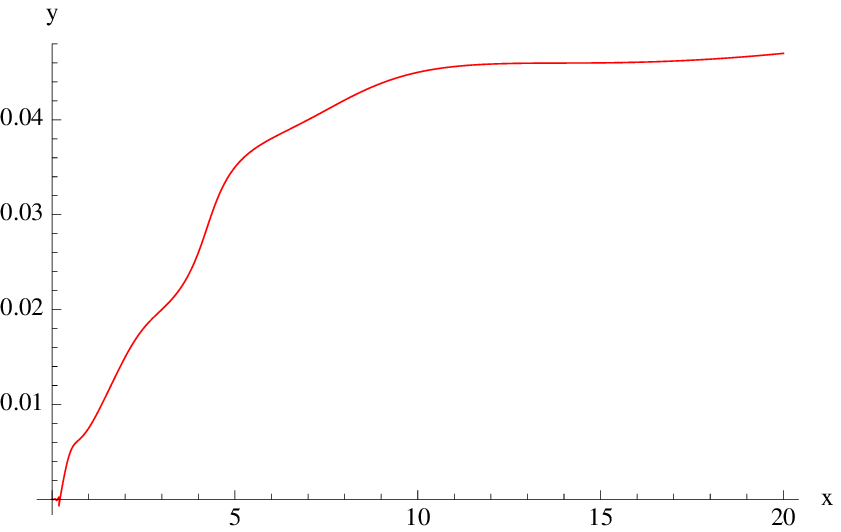} 
\end{center}
\caption{$\Deltagap$ in the anti-nodal direction as a function of $g$  (hole-doping $h=0.15$).}  
\vspace{-2mm}
\label{Gapofg} 
\end{figure} 

Numerically,   we found no solutions to the eqn.  (\ref{gapeqn})  in the nodal direction.    
In fact,   solutions only exist in a narrow direction around the anti-nodal region.   
Since the anti-nodal direction is along a 1-dimensional chain of  lattice sites,   this suggests
that the effect we are describing  is closely tied to the properties of the 1-dimensional chains 
 studied in the appendix.    Although
this may help to explain the d-wave nature of the gap in the cuprates,   it by no means establishes it.  
The potential $\Vex$ is invariant under $90^\circ$ rotations of the Brillouin zone,   as is the hamiltonian.
A d-wave gap,  by definition,  spontaneously  breaks this symmetry,  i.e. it alternates in sign under such a rotation, thus
one cannot deduce a d-wave gap from the poles in $\Vex$ alone.   
In order to study this further,  one needs a BCS-like gap equation built upon the effective 
interaction $\Vex$,  or equations analagous to those in \cite{Raghu},  
 which is beyond the scope of this paper.    We point out that   It is known that a BCS gap equation based on a potential
 $V( \kvec, \kvec')$ which is invariant  $90^\circ$ rotations has both s and d-wave solutions\cite{Kotliar}.

\section{Conclusions}

In summary,  we identified the possibility of Cooper pairing instabilities which arise 
as poles in certain Green functions in the two dimensional   Hubbard model,   in a manner analagous  to the BCS theory.   
The only parameter in the calculation is the value of the coupling
$U/t = g$ at the reference temperature $T_0 /t = 0.001$,  which we took to be $g=10$.  
Reasonable magnitudes for the gap in the anti-nodal direction  were found,  $\Deltagap/t  < 0.08$, 
in the region of hole doping $0.03 < h < 0.24$,  
without introducing an explicit cut-off.     

In the BCS theory,  the analogous poles are `resolved' by the BCS gap equation.   
The latter  has not been developed in the present case,   thus this should be investigated
as a next step in exploring the consequences of the pairing mechanism identified  in this work.  
Since it is not clear that the usual BCS gap equation with the effective pair potential studied in 
this paper is valid,   this has not been pursued here.   
Certainly the very existence of these poles imply that the effective pairing potential can change sign
thereby becoming attractive,  thus non-zero solutions to the proper gap equation should exist.

\section{Acknowledgments}

We wish to thank Neil Ashcroft and Erich Mueller for discussions.   
    This work is supported by the National Science Foundation
under grant number  NSF-PHY-0757868.

\section{Appendix:   Chains:   The loop integrals in one dimension.}

In this appendix we study the one-dimensional version of the integrals for $\Lex$, i.e. 
eq. (\ref{Lexdef})  with $\int d^2 \pvec /(2\pi)^2 \to  \int d\pvec /2\pi $  where $\pvec$ is now
a one-dimensional vector,  and $\omega_\kvec = -2 \cos \kvec $.   
  In the limit of zero temperature,  the occupation number
$f$ becomes a step function,  and the integral can be performed analytically.     Namely,  
\beq
\label{A1}
\Lex =  L_+  -  L_-  
\eeq
where
\beq
\label{A2} 
L_+  = \inv{2\pi}  \int_{|\pvec| \leq \kvec_F (\mu) }     \(  
\frac{d\pvec}{\Deltaxi + \xi_\pvec - \xi_{\pvec + \qvec}} \) ,   ~~~~~
L_-   = \inv{2\pi}   \int_{|\pvec+ \qvec | \leq \kvec_F (\mu) }   \(  
\frac{d\pvec}{\Deltaxi + \xi_\pvec - \xi_{\pvec + \qvec}} \)
\eeq
where $\kvec_F (\mu)  = \arccos (-\mu/2)$ is the Fermi momentum.   
In the region of small $\qvec$ that we are interested in,  the appropriate branch of 
 the above integrals are given
in terms of the functions
\beq
\label{Idef} 
I( \pvec )  =  \inv{\pi  \sqrt{\Deltaxi^2 - 8 ( 1- \cos \qvec )}}  
\arctan \(   
\frac{ (\Deltaxi  + 2 - 2 \cos \qvec ) \tan (\pvec/2) -2 \sin \qvec }
{\sqrt{ \Deltaxi^2 - 8 (1-\cos \qvec)}}
\), 
\eeq
as follows:
\beq
\label{A3}
L_+  =  I ( \kvec_F (\mu) )  -  I (-\kvec_F (\mu) ),  ~~~~~  
L_-  =  I (\kvec_F (\mu) - \qvec )  -  I(-\kvec_F (\mu) -\qvec )
\eeq
In the limit of zero temperature,   $\Lex$ is  then  only a  function of $\qvec,  \Deltaxi$ 
and the chemical potential $\mu$, by using the identities
\beq
\label{A4}
\tan( \kvec_F (\mu)/2)  = \sqrt{  \frac{ 2+\mu}{2-\mu} }, ~~~~
\tan ( (\kvec_F (\mu) -\qvec)/2)  =   \frac{   \sqrt{2+\mu}  -  \sqrt{2-\mu} \tan(\qvec/2) }
{\sqrt{2-\mu}  + \sqrt{2+\mu} \tan (\qvec/2) }
\eeq

Using the above expressions, one can show that $\Lex$ can indeed be positive,  which is
required for pole singularities in $\Veffex$.   
For  instance,  for small $q/\Deltaxi$,   $\Lex$ has the following asymptotic form: 
\beq
\label{Lexsmallq}
\Lex (q, \Deltaxi, \mu)    \approx 
\frac{ q^2 \sqrt{4-\mu^2}}{ \Deltaxi^2  \pi }
\( 
1 + \frac{q^2}{\Deltaxi^2}  \(  4 - \mu^2 - \Deltaxi^2/12 \)  
\)
\eeq

\begin{figure}[htb] 
\begin{center}
\hspace{-15mm} 
\psfrag{x}{$q$}
\psfrag{y}{$\Lex$}
\psfrag{qc}{$q_F$}
\includegraphics[width=10cm]{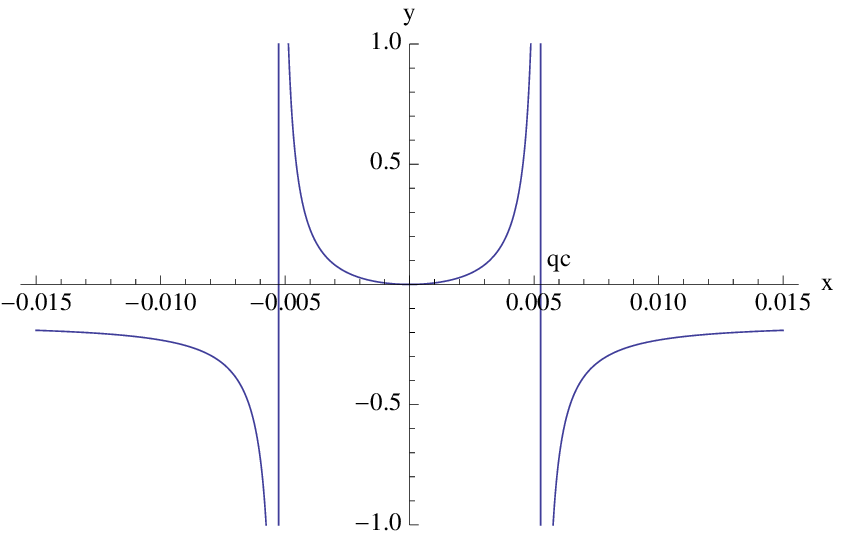} 
\end{center}
\caption{The loop integral $\Lex $  as a function of $q$ for for $\Deltaxi =0.01$ and  hole doping $h=0.20$.}  
\vspace{-2mm}
\label{qc} 
\end{figure}

Whereas $\qvec$ and $\Deltaxi$ were treated as independent in the above formulas,
physically we are interested in the situation where $\qvec = \kvec' - \kvec$ and $\Deltaxi = \xi_{\kvec'} - \xi_\kvec$,  referred to as `on-shell' above.     
In this case there is a divergence in $\Lex$ which arises from the pole in the integrand when $\pvec = \kvec$.
Since we are interested in $\kvec, \kvec'$ near the Fermi surface,  let us be more specific and
let $\kvec = \kvec_F (\mu) $  be right on the Fermi surface,  and $\kvec'$ be slightly off of it,  for a small $\xi_{\kvec'}$.
   Here $\Deltaxi = \xi_{\kvec'} $ and the singularity occurs at 
$\qvec_F \equiv \arccos(-(\xi_{\kvec'} + \mu)/2) - \kvec_F (\mu)$.   For $\xi_{\kvec'} =0.01$ and $\mu = -0.618$,  
which corresponds to hole doping $h=0.2$,  the singularity is at $\qvec_F = \pm  0.0053$, which is shown
in Figure \ref{qc}. 
  Analytically,  this divergence  arises as $\arctan (i)$ in the above formulas.

The $\pvec$-integral for $\Lex$ should thus be understood as the Cauchy principal value (PV), i.e.
$\int d\pvec \to  \lim_{\delta \to 0}  \(  \int_{-\pi}^{\kvec_F - \delta} d\pvec  +  \int_{\kvec_F + \delta}^\pi d \pvec \)$.  
For finite temperature $T$,  one can check numerically that this PV integral is  well-defined and finite.   
However it is important to note that $\Lex$ continues to be divergent as $T\to 0$,  which the above
analytic formulas demonstrate.   The reason is that in the limit of zero $T$,  $f$ is a step function
and the $\int_{\kvec_F + \delta}^\pi  d\pvec$ piece of the PV is absent,  leading it to be ill-defined.
This is  interpreted as an infra-red divergence that needs to be regulated by
a finite $T$,  however small.

\begin{figure}[htb] 
\begin{center}
\hspace{-15mm} 
\psfrag{x}{$h$}
\psfrag{y}{$\Deltagap$}
\includegraphics[width=7cm]{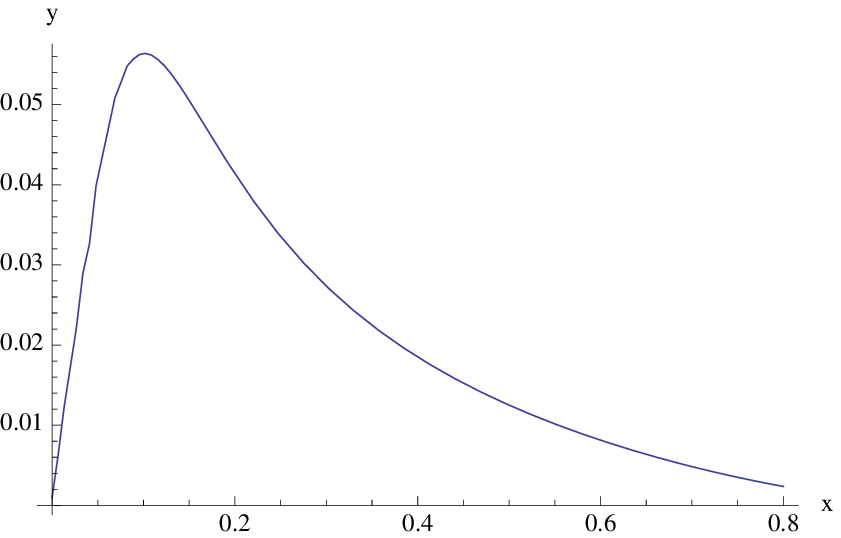} 
\end{center}
\caption{The $\Deltagap$ pole in $\Veffex$  as a function of  hole doping $h$
for bare coupling $g=1$.  }  
\vspace{-2mm}
\label{Gapofh1D} 
\end{figure}

\begin{figure}[htb] 
\begin{center}
\hspace{-15mm} 
\psfrag{x}{$g$}
\psfrag{y}{$\Deltagap$}
\includegraphics[width=7cm]{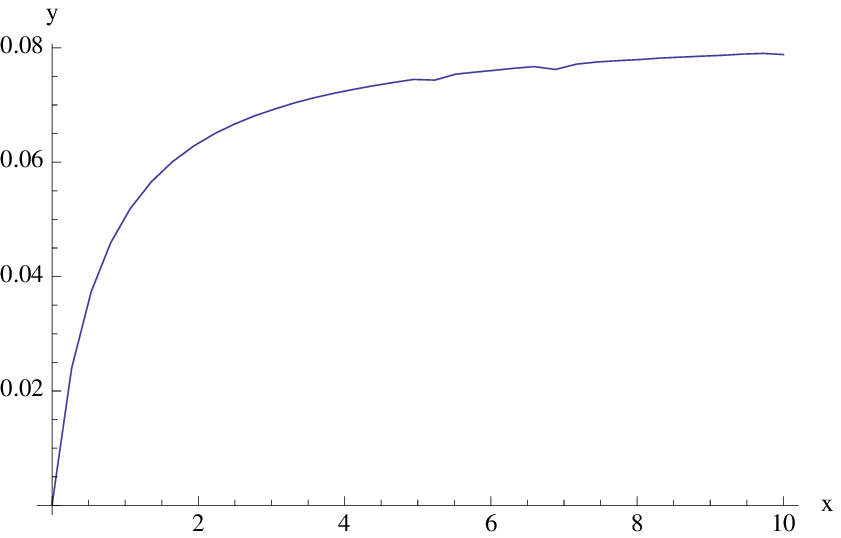} 
\end{center}
\caption{The $\Deltagap$ pole in $\Veffex$  as a function of coupling $g$ for  hole doping $h=0.15$.}  
\vspace{-2mm}
\label{Deltaofg} 
\end{figure}

Let now turn to the 1-d analogs of the gap $\Deltagap$ defined in section III.   In this 1d case, 
because of the divergence as $T\to 0$,  it makes sense to define $\Deltagap$ as solutions to 
eqn.  (\ref{gapeqn}) with $T=\Deltagap$,  i.e.  to determine $\Deltagap$ at a temperature comparable to it. 
The result is shown in Figure \ref{Gapofh1D},  where $\gsc$ was taken to be the screened coupling for a bare
$g=1$.       In Figure \ref{Deltaofg} we plot $\Deltagap$ as a function
of the bare coupling $g$;  here,  contrary to the 2d case,  there is no minimal value of $g$ required
for the existence of solutions.     On the other hand the $\Deltagap$ saturates with increasing $g$ 
as in the two-dimensional case.

\end{document}